\documentclass[12pt]{article}
\usepackage{axodraw,bbold}

\catcode`@=12
\begin{document}
\thispagestyle{empty}
\begin{flushright} 
UCRHEP-T413\\ 
June 2006\
\end{flushright}
\vspace{0.5in}
\begin{center}
{\LARGE	\bf Lepton family symmetries\\ for neutrino masses and mixing\\}
\vspace{1.5in}
{\bf Ernest Ma\\}
\vspace{0.2in}
{\sl Physics Department, University of California, Riverside, 
California 92521\\}
\vspace{1.5in}
\end{center}
\begin{abstract}\
I review some of the recent progress (up to December 2005) in applying
non-Abelian dicrete symmetries to the family structure of leptons, with
particular emphasis on the tribimaximal mixing {\it ansatz} of Harrison,
Perkins, and Scott.
\end{abstract}
\vspace{1.0in}
\centerline{---~~{\it Talk at WHEPP-9}~~---}
\newpage
\baselineskip 24pt

\section{Some Basics}

Using present data from neutrino oscillations, the $3 \times 3$ neutrino
mixing matrix is largely determined, together with the two mass-squared
differences \cite{data}.  In the Standard Model of particle interactions,
there are 3 lepton families.  The charged-lepton mass matrix linking
left-handed $(e, \mu, \tau)$ to their right-handed counterparts is in
general arbitrary, but may always be diagonalized by 2 unitary
transformations:
\begin{equation}
{\cal M}_l = U^l_L \pmatrix{m_e & 0 & 0 \cr 0 & m_\mu & 0 \cr 0 & 0 & m_\tau}
(U^l_R)^\dagger.
\end{equation}
Similarly, the neutrino mass matrix may also be diagonalized by 2 unitary
transformations if it is Dirac:
\begin{equation}
{\cal M}^D_\nu = U^\nu_L \pmatrix{m_1 & 0 & 0 \cr 0 & m_2 & 0 \cr 0 & 0 &
m_3} (U^\nu_R)^\dagger,
\end{equation}
or by just 1 unitary transformation if it is Majorana:
\begin{equation}
{\cal M}^M_\nu = U^\nu_L \pmatrix{m_1 & 0 & 0 \cr 0 & m_2 & 0 \cr 0 & 0 &
m_3} (U^\nu_L)^T.
\end{equation}
Notice that whereas the charged leptons have individual names, the
neutrinos are only labeled as $1,2,3$, waiting to be named.
The observed neutrino mixing matrix is the mismatch between
$U^l_L$ and $U^\nu_L$, i.e.
\begin{eqnarray}
U_{l\nu} = (U^l_L)^\dagger U^\nu_L \simeq \pmatrix{0.83 & 0.56 & <0.2
\cr -0.39 & 0.59 & -0.71 \cr -0.39 & 0.59 & 0.71} \simeq \pmatrix{\sqrt{2/3}
& 1/\sqrt{3} & 0 \cr -1/\sqrt{6} & 1/\sqrt{3} & -1/\sqrt{2} \cr -1/\sqrt{6}
& 1/\sqrt{3} & 1/\sqrt{2}}.
\end{eqnarray}
This approximate pattern has been dubbed tribimaximal by Harrison, Perkins,
and Scott \cite{hps}.  Notice that the 3 vertical columns are evocative
of the mesons $(\eta_8,\eta_1,\pi^0)$ in their $SU(3)$ decompositions.\\

\noindent How can the HPS form of $U_{l\nu}$ be derived from a symmetry? The
difficulty comes from the fact that any symmetry defined in the basis
$(\nu_e,\nu_\mu,\nu_\tau)$ is automatically applicable to $(e,\mu,\tau)$
in the complete Lagrangian.  To do so, usually one assumes the canonical
seesaw mechanism and studies the Majorana neutrino mass matrix
\begin{equation}
{\cal M}_\nu = -{\cal M}^D_\nu {\cal M}_N^{-1} ({\cal M}^D_\nu)^T
\end{equation}
in the basis where ${\cal M}_l$ is diagonal; but the symmetry apparent
in ${\cal M}_\nu$ is often incompatible with a diagonal ${\cal M}_l$ with
3 very different eigenvalues.\\

\noindent Consider just 2 families. Suppose we want maximal 
$\nu_\mu-\nu_\tau$
mixing, then we should have
\begin{equation}
{\cal M}_\nu = \pmatrix{a & b \cr b & a} = {1 \over \sqrt2} \pmatrix{1 & -1
\cr 1 & 1} \pmatrix{a+b & 0 \cr 0 & a-b} {1 \over \sqrt2} \pmatrix{1 & 1 \cr
-1 & 1}.
\end{equation}
This seems to require the exchange symmetry $\nu_\mu \leftrightarrow
\nu_\tau$, but since $(\nu_\mu,\mu)$ and $(\nu_\tau,\tau)$ are $SU(2)_L$
doublets, we must also have $\mu \leftrightarrow \tau$ exchange.
Nevertheless, we still have the option of assigning $\mu^c$ and $\tau^c$.
If $\mu^c \leftrightarrow \tau^c$ exchange is also assumed, then
\begin{equation}
{\cal M}_l = \pmatrix{A & B \cr B & A} = {1 \over \sqrt2} \pmatrix{1 & -1
\cr 1 & 1} \pmatrix{A+B & 0 \cr 0 & A-B} {1 \over \sqrt2} \pmatrix{1 & 1 \cr
-1 & 1}.
\end{equation}
Hence $U_{l\nu} = (U^l_L)^\dagger U^\nu_L = 1$ and there is no mixing.
If $\mu^c$ and $\tau^c$ do not transform under this exchange, then
\begin{equation}
{\cal M}_l = \pmatrix{A & B \cr A & B} = {1 \over \sqrt2} \pmatrix{1 & -1
\cr 1 & 1} \pmatrix{\sqrt{2(A^2+B^2)} & 0 \cr 0 & 0} \pmatrix{c & s \cr
-s & c},
\end{equation}
where $c=A/\sqrt{A^2+B^2}$, $s=B/\sqrt{A^2+B^2}$.  Again $U_{l\nu} =
(U^l_L)^\dagger U^\nu_L = 1$.\\

\noindent Obviously a more sophisticated approach is needed.  To that end,
I will list some non-Abelian discrete symmetries based on geometric solids,
in anticipation that some of them will be useful for realizing the HPS
{\it ansatz}.  I then focus on the tetrahedral group $A_4$
and show how the charged-lepton and neutrino mass matrices may be
constrained, followed by a catalog of recent models, with one detailed
example.  I will also discuss the symmetry $S_4$ with another example
and mention briefly the symmetry $B_4$.  These examples show
how exact and approximate tribimaximal mixing may be obtained.

\section{Some Discrete Symmetries}

\noindent The five perfect geometric solids were known to the
ancient Greeks.  In order to match them to the 4 elements (fire, air,
earth, and water) already known, Plato invented a fifth (quintessence)
as that which pervades the cosmos and presumably holds it together.
Since a cube (hexahedron) may be embedded inside an octahedron and vice versa,
the two must have the same group structure and are thus dual to each other.
The same holds for the icosahedron and dodecahedron.  The tetrahedron is
self-dual. Compare this first theory of everything to today's contender,
i.e. string theory.  (A) There are 5 consistent string theories in 10
dimensions. (B) Type I is dual to Heterotic $SO(32)$, Type IIA
is dual to Heterotic $E_8 \times E_8$, and Type IIB is self-dual.

\begin{table}[htb]
\centering
\caption{Perfect geometric solids in 3 dimensions.}
\begin{tabular}{cccccc}
\hline
solid&faces&vertices&Plato&Hindu&group\\
\hline
tetrahedron&4&4&fire&Agni&$A_4$\\
octahedron&8&6&air&Vayu&$S_4$\\
cube&6&8&earth&Prithvi&$S_4$\\
icosahedron&20&12&water&Jal&$A_5$\\
dodecahedron&12&20&quintessence&Akasha&$A_5$\\
\hline
\end{tabular}
\end{table}

\noindent Plato inferred the existence of the fifth element (quintessence) 
from the mismatch of the 4 known elements with the 5 perfect geometric solids.
In much the same way, Glashow, Iliopoulos, and Maiani inferred the
existence of the 4th quark (charm) from the mismatch of the 3 known
quarks (up, down, strange) with the 2 charged-current doublets
$(u, d \cos \theta_C + s \sin \theta_C)$ and $(?, -d \sin \theta_C +
s \cos \theta_C)$.\\

\noindent {\it Question}: What sequence has $\infty$, 5, 6 , 3, 3, 3,...?

\noindent {\it Answer}: Perfect geometric solids in 2, 3, 4, 5, 6, 7,...
dimensions.  In 2 dimensions, they are the regular polygons. In 4 
dimensions, they are:

\begin{table}[htb]
\centering
\caption{Perfect geometric solids in 4 dimensions.}
\begin{tabular}{cccc}
\hline
solid&composition&faces&vertices\\
\hline
4-simplex&tetrahedron&5&5\\
4-crosspolytope&tetrahedron&16&8\\
4-cube&cube&8&16\\
600-cell&tetrahedron&600&120\\
120-cell&dodecahedron&120&600\\
24-cell&octahedron&24&24\\
\hline
\end{tabular}
\end{table}

\noindent In 5 and more dimensions, only the first 3 types of solids 
continue.

\section{Tetrahedral Symmetry $A_4$}

For 3 families, we should look for a group with a \underline{3}
representation, the simplest of which is $A_4$, the group of the even
permutation of 4 objects, which is also the symmetry group of the
tetrahedron.
\begin{table}[htb]
\centering
\caption{Character table of $A_4$.}
\begin{tabular}{ccccccc}
\hline
class&$n$&$h$&$\chi_1$&$\chi_{1'}$&$\chi_{1''}$&$\chi_3$\\
\hline
$C_1$&1&1&1&1&1&3\\
$C_2$&4&3&1&$\omega$&$\omega^2$&0\\
$C_3$&4&3&1&$\omega^2$&$\omega$&0\\
$C_4$&3&2&1&1&1&--1\\
\hline
\end{tabular}
\end{table}

\noindent In the above, $\omega = \exp (2\pi i/ 3) = -(1/2) + i
(\sqrt 3/2)$, and the fundamental multiplication rule of $A_4$ is
\begin{eqnarray}
\underline{3} \times \underline{3} &=& \underline{1}(11+22+33) +
\underline{1}'(11+\omega^222+\omega33) + \underline{1}''
(11+\omega22+\omega^233) \nonumber \\ &+& \underline{3}(23,31,12) +
\underline{3}(32,13,21).
\end{eqnarray}
Note that $\underline{3} \times \underline{3} \times \underline{3} =
\underline{1}$ is possible in $A_4$, i.e. $a_1 b_2 c_3 +$ permutations,
and $\underline{2} \times \underline{2} \times \underline{2} = \underline{1}$
is possible in $S_3$, i.e. $a_1 b_1 c_1 + a_2 b_2 c_2$.\\

\noindent Consider $(\nu_i,l_i) \sim \underline{3}$ under $A_4$, then 
${\cal M}_\nu$ is of the form
\begin{equation}
{\cal M}_\nu = \pmatrix{a+b+c & f & e \cr f & a + b\omega + c\omega^2 & d \cr
e & d & a+b\omega^2+c\omega},
\end{equation}
where $a$ comes from $\underline{1}$, $b$ from $\underline{1}'$, 
$c$ from $\underline{1}''$, and $(d,e,f)$ from $\underline{3}$.  In this 
basis, ${\cal M}_l$ is generally not diagonal, but under $A_4$, there 
are two interesting cases.\\

\noindent {\bf (I)} Let $l^c_i \sim \underline{1}, \underline{1}',
\underline{1}''$, then with $(\phi_i^0,\phi_i^-) \sim \underline{3}$,
\begin{eqnarray}
{\cal M}_l &=& \pmatrix{h_1v_1 & h_2v_1 & h_3v_1 \cr h_1 v_2 & h_2 v_2 \omega
& h_3 v_2 \omega^2 \cr h_1 v_3 & h_2 v_3 \omega^2 & h_3 v_3 \omega} 
\nonumber \\ &=&
{1 \over \sqrt 3} \pmatrix{1 & 1 & 1 \cr 1 & \omega & \omega^2
\cr 1 & \omega^2 & \omega} \pmatrix{h_1 & 0 & 0 \cr 0 & h_2 & 0 \cr 0 & 0
& h_3} \sqrt{3} v,
\end{eqnarray}
for $v_1=v_2=v_3=v$.\\

\noindent {\bf (II)} Let $l^c_i \sim \underline{3}$, but $(\phi_i^0,\phi_i^-) 
\sim \underline{1}, \underline{1}', \underline{1}''$, then ${\cal M}_l$ is 
diagonal with
\begin{equation}
\pmatrix{m_e \cr m_\mu \cr m_\tau} = \pmatrix{1 & 1 & 1 \cr 1 & \omega
& \omega^2 \cr 1 & \omega^2 & \omega} \pmatrix{h_1v_1 \cr h_2v_2 \cr h_3v_3}.
\end{equation}
In either case, it solves the fundamental theoretical problem of having a 
symmetry for the neutrino mass matrix even though the charged-lepton mass 
matrix has three totally different eigenvalues.\\

\noindent To proceed further, the 6 parameters of ${\cal M}_\nu$ must be 
restricted, from which $U_{l\nu}$ may be obtained:
\begin{equation}
U^\dagger_L {\cal M}_\nu U^*_L = {\cal M}^{(e,\mu,\tau)}_\nu = 
U_{l\nu} \pmatrix{m_1 & 0 & 0 \cr 0 & m_2 & 0 \cr 0 & 0 & m_3} (U_{l\nu})^T,
\end{equation}
where
\begin{equation}
{\bf (I)}: ~~ U_L = {1 \over \sqrt 3} \pmatrix{1 & 1 & 1 \cr 1 & \omega & 
\omega^2 \cr 1 & \omega^2 & \omega}, ~~~~~ {\bf (II)}: ~~ U_L = 1.
\end{equation}

\section{Neutrino Mass Models}

Using {\bf (I)}, the first two proposed $A_4$ models start with only 
$a \neq 0$, yielding thus 3 degenerate neutrino masses.  In Ma and 
Rajasekaran \cite{mr01}, the degeneracy is broken sofly by $N_iN_j$ terms, 
allowing $b,c,d,e,f$ to be nonzero.  In Babu, Ma, and Valle \cite{bmv03}, 
the degeneracy is broken radiatively through flavor-changing supersymmetric 
scalar lepton mass terms.  In both cases, $\theta_{23} \simeq \pi/4$ is 
predicted.  In BMV03, maximal CP violation in $U_{l\nu}$ is also predicted. 
Consider the case $b=c$ and $e=f=0$ \cite{m04}, then
\begin{eqnarray}
{\cal M}^{(e,\mu,\tau)}_\nu &=& \pmatrix{a+2d/3 & b-d/3 & b-d/3 \cr b-d/3 &
b+2d/3 & a-d/3 \cr b-d/3 & a-d/3 & b+2d/3} \nonumber \\
&=& a \pmatrix{1 & 0 & 0 \cr 0 & 0 & 1\cr 0 & 1 & 0} + b \pmatrix{0 & 1 & 1 
\cr 1 & 1 & 0 \cr 1 & 0 & 1} + {d \over 3} \pmatrix{2 & -1 & -1 
\cr -1 & 2 & -1\cr -1 & -1 & 2} \nonumber \\ 
&=& U_{l\nu} \pmatrix{a-b+d & 0 & 0 \cr 0 & a+2b & 0 \cr 0 & 0 & -a+b+d} 
(U_{l\nu})^T,
\end{eqnarray}
where
\begin{equation}
U_{l\nu} = \pmatrix{\sqrt{2/3}
& 1/\sqrt{3} & 0 \cr -1/\sqrt{6} & 1/\sqrt{3} & -1/\sqrt{2} \cr -1/\sqrt{6}
& 1/\sqrt{3} & 1/\sqrt{2}},
\end{equation}
i.e. tribimaximal mixing would be achieved.  However, although $b \neq c$ 
would allow $U_{e3} \neq 0$, the assumption $e=f=0$ and the bound $|U_{e3}| 
< 0.16$ together imply $0.5 < \tan^2 \theta_{12} < 0.52$, whereas 
experimentally, $\tan^2 \theta_{12} = 0.45 \pm 0.05$.\\

\noindent Other models based on {\bf (I)} with $d \neq 0$ and $e=f=0$ 
include AF05-1/2 \cite{af05-1,af05-2} with $b=c=0$; M05-1 \cite{m05-1} 
with $a=0$ and assuming $b=c$; BH05 \cite{bh05} with $b=c$ and $d^2=3b(b-a)$; 
Z05 \cite{z05} assuming $b=c$, and M05-5 \cite{m05-5} with $b=c$.
They are summarized in Table 4.
\begin{table}[htb]
\centering
\caption{Particle content of models based on {\bf (I)}.}
\begin{tabular}{cccccc}
\hline
$A_4${\bf (I)}&$\phi^+,\phi^0$&$N$&$\xi^{++},\xi^+,\xi^0$&$\chi^0$&SUSY\\
\hline
MR01&1,3&3&---&---&no\\
BMV03&1,1&3&---&3&yes\\
M04&3&---&$1,1',1'',3$&---&no\\
AF05-1&1,1&---&---&1&1,3,3\\
M05-1&3&---&$1',1'',3$&---&no\\
BH05&1,1&3&---&1,1,3,3&yes\\
Z05&3&---&---&---&no\\
M05-5&1,1&3&---&3&yes\\
AF05-2&1,1&---&---&1,1,1,3,3,3,3&yes\\
\hline
\end{tabular}
\end{table}

\noindent Consider as an example M05-5 \cite{m05-5}.  Here
\begin{equation}
{\cal M}^D_\nu = U^\dagger_L \pmatrix{m_D & 0 & 0 \cr 0 & m_D & 0 
\cr 0 & 0 & m_D}, ~~~ {\cal M}_N = \pmatrix{A & 0 & 0 \cr 0 & B & C \cr 
0 & C & B}
\end{equation}
imply $e=f=0$ and $b=c$.  To obtain this ${\cal M}_N$, consider the 
superpotential
\begin{eqnarray}
W &=& {1 \over 2} m_N (N_1^2+N_2^2+N_3^2) + fN_1N_2N_3 \nonumber \\
&+& {\lambda_1 \over 4 M_{PL}} (N_1^4+N_2^4+N_3^4) + {\lambda_2 \over 
2 M_{Pl}} (N_2^2N_3^2+N_3^2N_1^2+N_1^2N_2^2),
\end{eqnarray}
and its resulting scalar potential
\begin{eqnarray}
V &=& |m_N N_1 + f N_2 N_3 + {\lambda_1 \over M_{Pl}} N_1^3 + {\lambda_2 
\over M_{Pl}} N_1(N_2^2+N_3^2)|^2 \nonumber \\ 
&+& |m_N N_2 + f N_3 N_1 + {\lambda_1 \over M_{Pl}} N_2^3 + {\lambda_2 
\over M_{Pl}} N_2(N_3^2+N_1^2)|^2 \nonumber \\ 
&+& |m_N N_3 + f N_1 N_2 + {\lambda_1 \over M_{Pl}} N_3^3 + {\lambda_2 
\over M_{Pl}} N_3(N_1^2+N_2^2)|^2.
\end{eqnarray}
The usual solution of $V=0$ is $\langle N_{1,2,3} \rangle = 0$, but the 
following is also possible:
\begin{equation}
\langle N_{2,3} \rangle = 0, ~~~ \langle N_1 \rangle^2 = {-m_N M_{Pl} \over 
\lambda_1},
\end{equation}
yielding the above form of ${\cal M}_N$ with
\begin{equation}
A = -2m_N, ~~~ B = (1-\lambda_2/\lambda_1)m_N, ~~~ C=f \langle N_1 \rangle.
\end{equation}
The soft term $-h\langle N_1 \rangle (\nu_1 \phi^0-l_1\phi^+)$ must also be 
added to allow $(\nu_1,l_1)$ and $(\phi^+,\phi^0)$ to remain massless at the 
seesaw scale.  The resulting theory is then protected below the seesaw 
scale by the usual $R$-parity of a supersymmetric theory.  Thus $A_4$ 
allows tribimaximal neutrino mixing to be generated automatically from the 
$N_{1,2,3}$ superfields themselves.  However, the neutrino mass eigenvalues 
are not predicted.\\

\noindent Consider next the assignments of case {\bf (II)}.  Here 
${\cal M}_\nu^{(e,\mu,\tau)}= {\cal M}_\nu$ already.  Let $d=e=f$, then
\begin{equation}
{\cal M}_\nu = \pmatrix{a+b+c & d & d \cr d & a+b\omega+c\omega^2 & d \cr
d & d & a+b\omega^2+c\omega}.
\end{equation}
Assume $b=c$ and rotate to the basis $[\nu_e,(\nu_\mu+\nu_\tau)/\sqrt 2,
(-\nu_\mu+\nu_\tau)/\sqrt 2]$, then
\begin{equation}
{\cal M}_\nu = \pmatrix{a+2b & \sqrt 2 d & 0 \cr \sqrt 2 d & a-b+d & 0 \cr
0 & 0 & a-b-d},
\end{equation}
i.e. maximal $\nu_\mu - \nu_\tau$ mixing and $U_{e3}=0$.  The solar mixing
angle is now given by $\tan 2 \theta_{12} = 2\sqrt 2 d/(d-3b)$.  For
$b << d$, $\tan 2 \theta_{12} \to 2\sqrt 2$, i.e. $\tan^2 \theta_{12} \to
1/2$, but $\Delta m^2_{sol} << \Delta m^2_{atm}$ implies $2a+b+d \to 0$, so
that $\Delta m^2_{atm} \to 6bd \to 0$ as well.  Therefore, $b \neq 0$ is
required, and $\tan^2 \theta_{12} \neq 1/2$, but should be close to it,
because $b=0$ enhances the symmetry of ${\cal M}_\nu$ from $Z_2$ to $S_3$.
Here $\tan^2 \theta_{12} < 1/2$ implies inverted ordering and
$\tan^2 \theta_{12} > 1/2$ implies normal ordering.\\

\noindent Models based on {\bf (II)} include CFM05 \cite{cfm05} with 
$3b = -ef/d-\omega^2fd/e-\omega de/f$ and $3c = -ef/d-\omega fd/e-\omega^2 
de/f$; M05-2 \cite{m05-2} with 2 complicated equalities; HMVV05 \cite{hmvv05} 
with $d=e=f$ and assuming $b=c$; and M05-3 \cite{m05-3} with $b=c$, $e=f$, 
and $(a+2b)d^2=(a-b)e^2$. They are summarized in Table 5.
\begin{table}[htb]
\centering
\caption{Particle content of models based on {\bf (II)}.}
\begin{tabular}{ccccc}
\hline
$A_4${\bf (II)}&$\phi^+,\phi^0$&$N$&$\xi^{++},\xi^+,\xi^0$&$\chi^0$\\
\hline
CFM05&$1,1',1''$&3&1&3\\
M05-2&1&3&3&$1,1',1'',3$\\
HMVV05&$1,1',1''$&---&$1,1',1'',3$&---\\
M05-3&$1,1',1'',3$&$1,1',1''$&---&---\\
\hline
\end{tabular}
\end{table}

\section{Permutation Symmetry $S_4$}

In the above application of $A_4$, approximate tribimaximal mixing involves
the {\it ad hoc} assumption $b=c$.  This problem is overcome by using $S_4$ in
a supersymmetric seesaw model proposed recently \cite{m05-4}, yielding the
result
\begin{equation}
{\cal M}_\nu = \pmatrix{a+2b & e & e \cr e & a-b & d \cr
e & d & a-b}.
\end{equation}
Here $b=0$ and $d=e$ are related limits.  The permutation group of 4
objects is $S_4$.  It contains both $S_3$ and $A_4$.  It is also the
symmetry group of the hexahedron (cube) and the octahedron.
\begin{table}[htb]
\centering
\caption{Character table of $S_4$.}
\begin{tabular}{cccccccc}
\hline
class&$n$&$h$&$\chi_1$&$\chi_{1'}$&$\chi_2$&$\chi_3$&$\chi_{3'}$\\
\hline
$C_1$&1&1&1&1&2&3&3\\
$C_2$&3&2&1&1&2&--1&--1\\
$C_3$&8&3&1&1&--1&0&0\\
$C_4$&6&4&1&--1&0&--1&1\\
$C_5$&6&2&1&--1&0&1&--1\\
\hline
\end{tabular}
\end{table}

\noindent The fundamental multiplication rules are
\begin{eqnarray}
\underline{3} \times \underline{3} &=& \underline{1}(11+22+33) +
\underline{2}(11+\omega^222+\omega33,11+\omega22+\omega^233) \nonumber \\
&+& \underline{3}(23+32,31+13,12+21) + \underline{3}'(23-32,31-13,12-21),\\
\underline{3}' \times \underline{3}' &=& \underline{1} +
\underline{2} + \underline{3}_S + \underline{3}'_A,\\
\underline{3} \times \underline{3}' &=& \underline{1}' +
\underline{2} + \underline{3}'_S + \underline{3}_A.
\end{eqnarray}
Note that both $\underline{3} \times \underline{3} \times \underline{3} =
\underline{1}$ and $\underline{2} \times \underline{2} \times \underline{2}
= \underline{1}$ are possible in $S_4$.
Let $(\nu_i,l_i),l^c_i,N_i \sim \underline{3}$ under $S_4$.  Assume singlet
superfields $\sigma_{1,2,3} \sim \underline{3}$ and $\zeta_{1,2} \sim
\underline{2}$, then
\begin{equation}
{\cal M}_N = \pmatrix{M_1 & h \langle \sigma_3 \rangle & h \langle \sigma_2 
\rangle \cr h \langle \sigma_3 \rangle & M_2 & h \langle \sigma_1 \rangle \cr 
h \langle \sigma_2 \rangle & h \langle \sigma_1 \rangle & M_3},
\end{equation}
where $M_1 = A+f(\langle \zeta_2 \rangle + \langle \zeta_1 \rangle)$, 
$M_2 = A + f(\langle \zeta_2 \rangle \omega + \langle \zeta_1 \rangle 
\omega^2)$, and $M_3 = A + f(\langle \zeta_2 \rangle \omega^2 + \langle 
\zeta_1 \rangle \omega)$.  The most general $S_4$-invariant superpotential 
of $\sigma$ and $\zeta$ is
given by
\begin{eqnarray}
W &=& M(\sigma_1 \sigma_1 + \sigma_2 \sigma_2 + \sigma_3 \sigma_3) +
\lambda \sigma_1 \sigma_2 \sigma_3 + m \zeta_1 \zeta_2 + \rho(\zeta_1
\zeta_1 \zeta_1 + \zeta_2 \zeta_2 \zeta_2) \nonumber \\
&+& \kappa[(\sigma_1 \sigma_1 + \sigma_2 \sigma_2 \omega + \sigma_3 \sigma_3
\omega^2) \zeta_2 + (\sigma_1 \sigma_1 + \sigma_2 \sigma_2 \omega^2 +
\sigma_3 \sigma_3 \omega) \zeta_1].
\end{eqnarray}
The resulting scalar potential has a minimum at $V=0$ (thus preserving
supersymmetry) only if $\langle \zeta_1 \rangle = \langle \zeta_2 \rangle$
and $\langle \sigma_2 \rangle = \langle \sigma_3 \rangle$, so that
\begin{equation}
{\cal M}_N = \pmatrix{A+2B & E & E \cr E & A-B & D \cr
E & D & A-B}.
\end{equation}
To obtain a diagonal ${\cal M}_l$, choose $\phi^l_{1,2,3} \sim
\underline{1} + \underline{2}$.  To obtain a Dirac ${\cal M}_{\nu N}$
proportional to the identity, choose $\phi^N_{1,2,3} \sim \underline{1}
+ \underline{2}$ as well, but with zero vacuum expectation value for
$\phi^N_{2,3}$.  This allows ${\cal M}_\nu$ to have the form of Eq.~(24),
and thus approximate tribimaximal mixing.

\section{$B_4$}

Exact tribimaximal mixing has also been obtained by Grimus and Lavoura 
\cite{gl05} using the Coxeter group $B_4$, which is the symmetry group of 
the hyperoctahedron (4-crosspolytope) with 384 elements.  Here ${\cal M}_l$ 
is diagonal with $(\nu_i,l_i)$, $l^c_i$, and $(\phi^+_i,\phi^0_i)$ 
belonging to 3 different 3-dimensional representations of $B_4$ with 
the property $a_1b_1c_1 + a_2b_2c_2 + a_3b_3c_3 = 1$.  The ${\cal M}_N$ 
of Eq.~(30) is then reduced by $D=E+3B$. 

\section{Some Remarks}

With the application of the non-Abelian discrete symmetry $A_4$,
a plausible theoretical understanding of the HPS form of the neutrino mixing
matrix has been achieved, i.e. $\tan^2 \theta_{23} = 1$, $\tan^2 \theta_{12}
= 1/2$, $\tan^2 \theta_{13} = 0$.\\

\noindent Another possibility is that $\tan^2 \theta_{12}$ is not 1/2, but
close to it. This has theoretical support in an alternative version of $A_4$,
but is much more natural in $S_4$.\\

\noindent In the future, this approach to lepton family symmetry should be
extended to include quarks, perhaps together in a consistent overall theory.

\section*{Acknowledgement} I thank Pankaj Agrawal and the other organizers 
of WHEPP-9 for their great hospitality and a stimulating workshop.  This 
work was supported in part by the U.~S.~Department of Energy under Grant 
No.~DE-FG03-94ER40837.

\end{document}